\documentclass[aps,pra,showpacs,notitlepage,floatfix, twocolumn,
superscriptaddress,nofootinbib]{revtex4-1}

\usepackage{amsmath,amsthm,amsfonts,amssymb,amscd}
\usepackage{graphicx}
\graphicspath{ {./Figures/} }
\usepackage[table]{xcolor}
\usepackage[colorlinks,linkcolor=red,anchorcolor=red,citecolor=blue,urlcolor=blue]{hyperref}
\usepackage[titletoc]{appendix}
\usepackage{braket}
\usepackage[normalem]{ulem}

\def\be{\begin{equation}}
\def\ee{\end{equation}}
\def\bea{\begin{equation}\begin{aligned}}
\def\eea{\end{aligned}\end{equation}}
\def \tr {\text{tr}}

\usepackage{color}

\definecolor{navy}{RGB}{81, 114, 233}

\begin{document}

	\title{Scrambling dynamics across a thermalization-localization quantum phase transition}
	\author{Subhayan Sahu}
	\affiliation{Condensed Matter Theory Center and Department of Physics, University of Maryland, College Park, MD 20742, USA}
	\author{Shenglong Xu}
	\affiliation{Condensed Matter Theory Center and Department of Physics, University of Maryland, College Park, MD 20742, USA}
	\author{Brian Swingle}
	\affiliation{Condensed Matter Theory Center, Maryland Center for Fundamental Physics, Joint Center for Quantum Information and Computer Science, and Department of Physics, University of Maryland, College Park, MD 20742, USA}

	\begin{abstract}
		We study quantum information scrambling, specifically the growth of Heisenberg operators, in large disordered spin chains using matrix product operator dynamics to scan across the thermalization-localization quantum phase transition. We observe ballistic operator growth for weak disorder, and a sharp transition to a phase with sub-ballistic operator spreading. The critical disorder strength for the ballistic to sub-ballistic transition is well below the many body localization phase transition, as determined from finite size scaling of energy eigenstate entanglement entropy in small chains. In contrast, we find that the transition from sub-ballistic to logarithmic behavior at the actual eigenstate localization transition is not resolved in our finite numerics. These data are discussed in the context of a universal form for the growing operator shape and substantiated with a simple phenomenological model of rare regions.
		
	\end{abstract}
	
	
	\maketitle
	

	
	It has long been known that disorder can slow or arrest quantum motion \cite{Anderson1958}, leading to a localized state. Recently it was understood that localization can survive even strong interactions, a phenomenon dubbed many-body localization (MBL) \cite{Basko2006,Oganesyan2007,Gornyi2005}. More precisely, there is a quantum phase transition in interacting systems from a thermalizing phase to a localized phase with increasing disorder. The phase and phase transition have been intensely studied (e.g., \cite{Pal2010,Znidaric2008,Bardarson2012,Serbyn2013,Bauer2013,Altman2015,Serbyn2013b,Vosk2014,Huse2014a,Swingle2013a,
		Agarwal2015,Agarwal2017,Luitz2015b,Nandkishore2015,Potter2015,Abanin2017}), and there is a proof, given plausible assumptions, of the existence of MBL in one-dimensional spin chains with local interactions \cite{Imbrie2016,Imbrie2016a}.
	
	In this work we are particularly concerned with the quantum phase transition (or transitions) that take a one-dimensional disordered system from a thermalizing phase to a localized phase  \cite{Huse2013,Chandran2013,Bauer2013,Pekker2014,Vosk2014,Kjall2014,Grover2014a,Bahri2015,Khemani2017a}. It is natural to study this phase transition via dynamics \cite{Znidaric2008,Bardarson2012,Serbyn2013,Altman2015}, because eigenstate based numerics are difficult to scale to large system sizes and because dynamical properties are accessible in experiments \cite{Rubio-Abadal2018,Luschen2017,Luschen2017a}. We study a dynamical quantity related to quantum information scrambling, the squared commutator \cite{Shenker2014,Hayden2007,Sekino2008,Hosur2016}. 
	
	Consider two local operators, $W$ and $V$, in a one-dimensional spin chain, separated by a distance $x$. The squared commutator probes the extent to which $V$ fails to commute with the time evolved Heisenberg operator $W(t)=e^{i H t} W e^{-i H t}$. It is defined as the expectation value of the absolute value squared of the commutator of the $W(t)$ and $V$,
	\begin{equation}
	C(x,t)= \langle [W(t),V]^{\dagger}[W(t),V] \rangle.
	\end{equation}
	It is closely related to the out of time ordered correlator (OTOC), $F(t)=\braket{W^\dagger(t)VW^\dagger(t)V}$. OTOCs are currently receiving attention as a diagnostic of quantum chaos~\cite{Larkin1969,Shenker2014,Kitaev2015,Maldacena2016}, including experimental proposals~\cite{Swingle2016,Zhu2016,Yao2016,YungerHalpern2017} and early experiments measuring OTOCs~\cite{Li2017,Garttner2017,Wei2018,Meier2017}. In fact, \cite{Wei2018} measured OTOCs to detect localization in NMR spin systems. 
	
	The squared commutator starts at zero for initially separated $W$ and $V$, and then grows as the operator $W(t)$ spreads and overlaps with the location of $V$. In the absence of disorder, $C(x,t)$ typically grows ballistically, leading to an emergent linear light cone with butterfly velocity $v_B$. On the other hand, disorder can severely arrest the growth of $C(x,t)$, a manifestation of localization. It has been argued that MBL is characterized by an extensive number of local integrals of motion \cite{Serbyn2013b,Vosk2014,Huse2014a,Swingle2013a}, leading to an emergent logarithmic light cone \cite{Burrell2007}. Similarly, it was recently shown that the disorder averaged $C(x,t)$ exhibits a logarithmic light cone with $v_B=0$ in the MBL phase~\cite{Swingle2017a,Huang2017a,Fan2017,He2017,Chen2016,Chen2017,Slagle2017}.
	
	In this letter we study operator dynamics across the entire thermal-to-MBL phase diagram, with a particular focus on the thermal side of the MBL eigenstate transition. This regime has attracted interest in the context of rare region effects which can slow down transport well before the MBL transition \cite{Agarwal2015,Agarwal2017,Znidaric2016,Khait2016}. One interesting question is whether the butterfly velocity survives arbitrarily weak disorder~\cite{Luitz2017,Nahum2017}. 
It is challenging, since, for example, strong disorder RG \cite{Slagle2017} applies only in the MBL phase and state-of-the-art exact diagonalization is still limited to small sizes~\cite{Luitz2017}. 
We use a recent t-DMRG based matrix product operator method to calculate dynamics of local Heisenberg operators \cite{Xu2018} (see also \cite{Lin2018,Khemani2018}) for larger system sizes ($\mathcal{O}(200)$ spins) and longer times than previously possible.
	
	\begin{figure}
		\includegraphics[width=\linewidth]{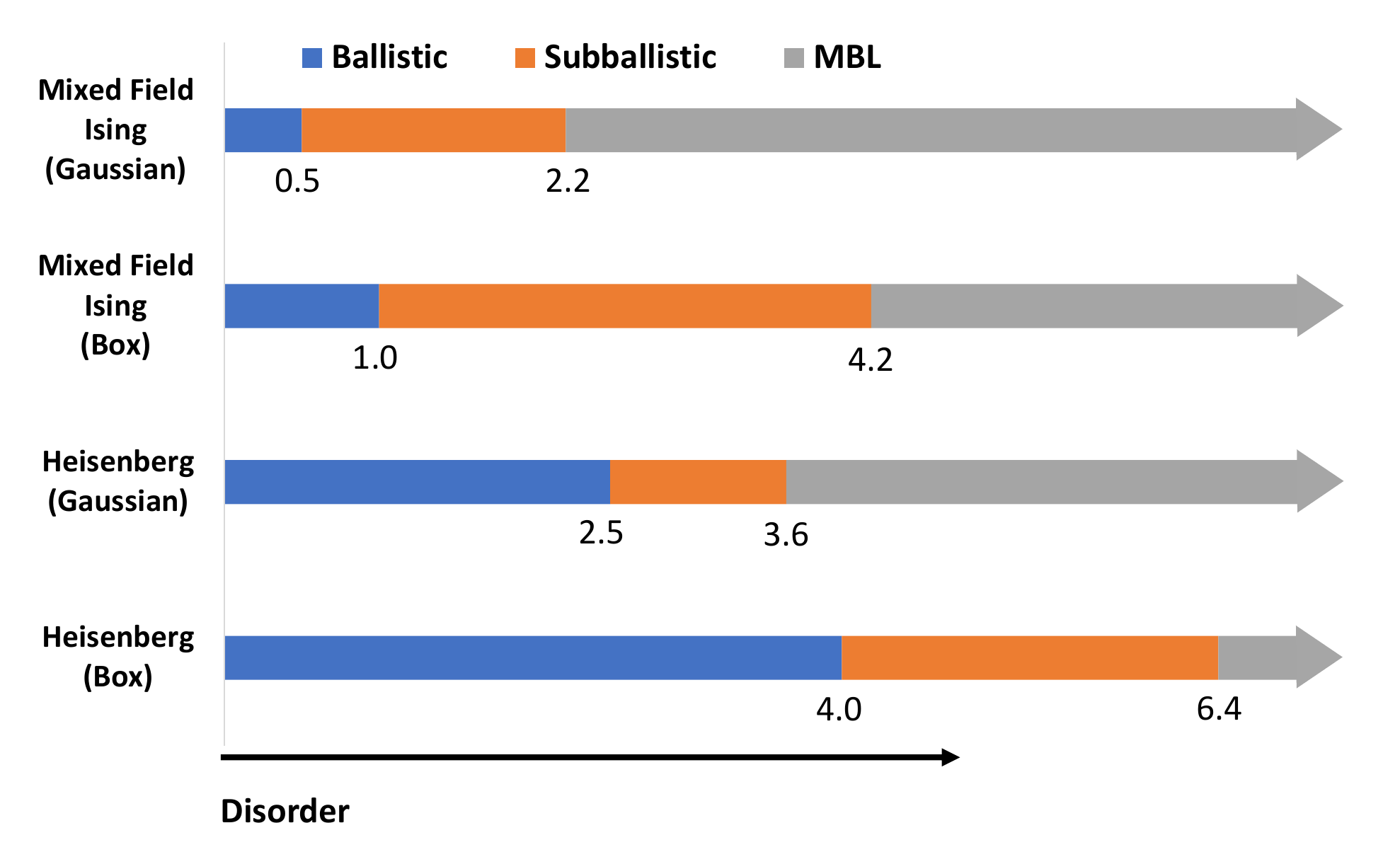}
		\caption{ Phase diagram of operator spreading in disordered interacting spin systems with different disorder models. The Heisenberg Hamiltonian is defined using Pauli operators instead of spin-1/2 operators, so the $W$ normalization is twice as large relative to the spin-1/2 convention.}	
		\label{Fig:phasediag}
	\end{figure}
	
	First, we observe a weak disorder phase with ballistic operator spreading $(v_B \neq 0)$ as well as a sharp transition to a sub-ballistic phase $(v_B = 0)$, at a disorder strength well below the putative MBL transition. This transition is characterized by a continuous vanishing of $v_B$ and an apparent divergence of the wavefront broadening. Second, we study the variability of operator growth from one disorder realization to another, which also characterize the ballistic to sub-ballistic transition independent of the fitting procedure. This is also a clear numerical demonstration of rare regions which is only possible because of the large system size. Observations from the variability of the scrambling data motivate a simple phenomenological model of rare regions, from which we analytically substantiate the presence of the ballistic phase. Together these numerical observations reveal a rich dynamical phase diagram for disordered spin models (Fig.~\ref{Fig:phasediag}). Comparing to previous studies, we find that the loss of ballistic operator spreading occurs at a larger disorder strength than the diffusive to sub-diffusive transition in spin transport, indicating at least four non-trivial dynamical regimes \cite{Agarwal2015,Agarwal2017,Znidaric2016,Khait2016,Nahum2017,Mendoza-Arenas2018}. 
	\par

	\textit{Model -- }For concreteness, we consider two one-dimensional spin chain models: \newline 1. Mixed field Ising model with $\sigma^z$ disorder
	\begin{equation} \label{eq:MFI}
	H = -J\sum_{r=1}^{L-1} Z_{r}Z_{r+1} - h_x\sum_{r=1}^{L}X_r - \sum_{r=1}^{L} h_{z,r} Z_r
	\end{equation}
	\newline 2. Heisenberg model with $\sigma^z$ disorder,
	\begin{equation} \label{eq:HMD}
	H = -J\sum_{r=1}^{L-1} \left(X_{r}X_{r+1}+Y_{r}Y_{r+1}+Z_{r}Z_{r+1}\right) - \sum_{r=1}^{L} h_{z,r} Z_r.
	\end{equation}
	Here $X_r, Y_r, Z_r$ are the local Pauli operators. For the mixed field Ising model, we choose the parameters $J=1$, $h_x=1.05$ and $\overline{h_{z,r}}=0.5$. For the Heisenberg model,  we choose the parameters $J=1$ and $\overline{h_{z,r}}=0$. For each spin chain we consider two different disorder probability distributions, box and Gaussian. For the box disorder, we draw the $h_{z,r}$ fields uniformly at random from the interval $[-W,W]$, with $W$ being the disorder strength. For Gaussian disorder, the $h_{z,r}$ fields are Gaussian random variables with standard deviation (SD) $W$. The parameters for the mixed field Ising model have been chosen so that the $W=0$ limit is strongly chaotic \cite{Xu2018}. The Heisenberg model with box disorder has been extensively studied for chains with $L\lesssim 30$ spins, and it has been shown that the thermal-MBL transition occurs at $W \gtrsim 7$ \cite{Luitz2015b}. We consider all these models to elucidate the robustness of the intermediate regime, and also to understand the role of disorder distribution on rare region effects. 
	
	\textit{Method -- }Our technique is a real-time tensor network method for operator dynamics~\cite{Xu2018}. Studying real-time quantum dynamics using tensor network methods, such as state-based TEBD or t-DMRG methods \cite{Vidal2004,Vidal2003,Daley2004,White2004,Bardarson2012,Serbyn2013}, is typically limited to early times, because the entanglement of the state is upper-bounded by $\log(\chi)$, where $\chi$ is the bond dimension of the matrix product state (MPS) \cite{Bardarson2012}. However, in a recent paper \cite{Xu2018}, some of us have shown that by going to the Heisenberg picture, one can reliably access a much wider space-time region using dynamics of matrix product operators (MPO) because of the entanglement structure of the Heisenberg operator. The complexity of the operator only builds up within the lightcone and is not essential for studying the dynamical property of the wavefront. As a result, the butterfly velocity and the broadening of the wavefront can be accurately extracted from TEBD simulation on Heisenberg operators in the matrix product form with modest bond dimension.
	
\begin{figure}
		\centering
		\includegraphics[width=\columnwidth]{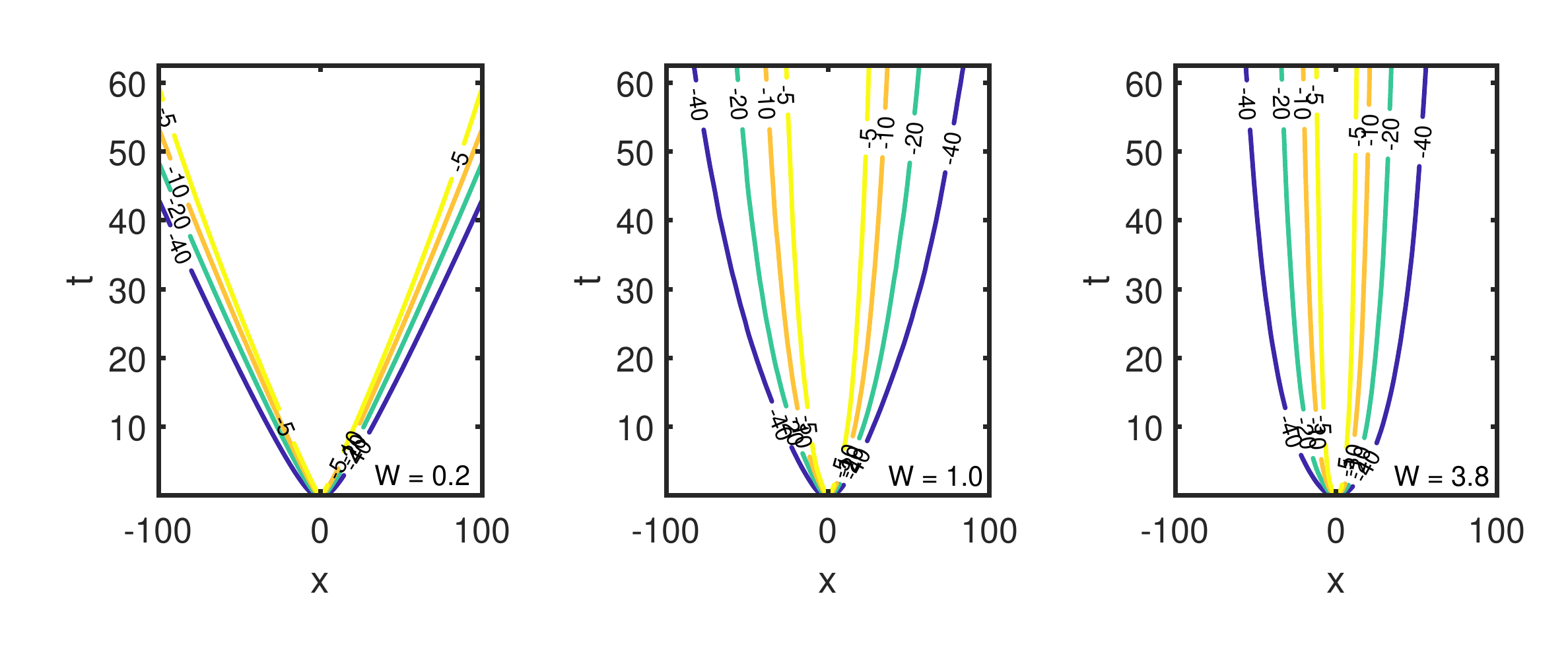}
		\caption{Plot of the contours of the averaged $\log(C)$, for the Mixed Field Ising model with Gaussian disorder. (averaged over $\sim 200$ disorder realizations, for three disorders, $W=0.2$ (ballistic), $W=1.0$ (intermediate) and $W=3.8$ (logarithmic). Bond dimension is 32. Convergence with bond dimension is discussed in the Supplemental Material. Fluctuations away from the disorder averaging are discussed in Fig.~\ref{Fig:variance} and in the corresponding section.)}
		\label{Fig:lightcone}
\end{figure}

	We simulated the squared commutator in the infinite temperature Gibbs ensemble, 
	\begin{equation}\label{eq:cdef}
	C(r-r^{\prime},t)=\frac{1}{2^{L}}\tr([X_r(t),X_{r^{\prime}}]^\dagger [X_r(t),X_{r^{\prime}}])
	\end{equation}
for spin chains of length $L = 201$ with maximal time of order $50-100$, in the units of $J^{-1}=1$. A small Trotter step of $\delta t = 0.0025$ is used to obtain high numerical precision. For each disorder, we consider around $200-500$ disorder realizations and average $\log(C)$ over the different realizations. This ensures that rare disorder realizations which could localize the operator growth are not overwhelmed by the ballistic samples during the averaging process. Fig.~\ref{Fig:lightcone} shows  light cone obtained from averaging $C(x,t)$  for different disorders, representing each phase in Fig. \ref{Fig:phasediag}. We discuss convergence of the numerical procedure in Sec I of S.M..
	
	\begin{figure}	
		\centering
		\includegraphics[width=0.7\columnwidth]{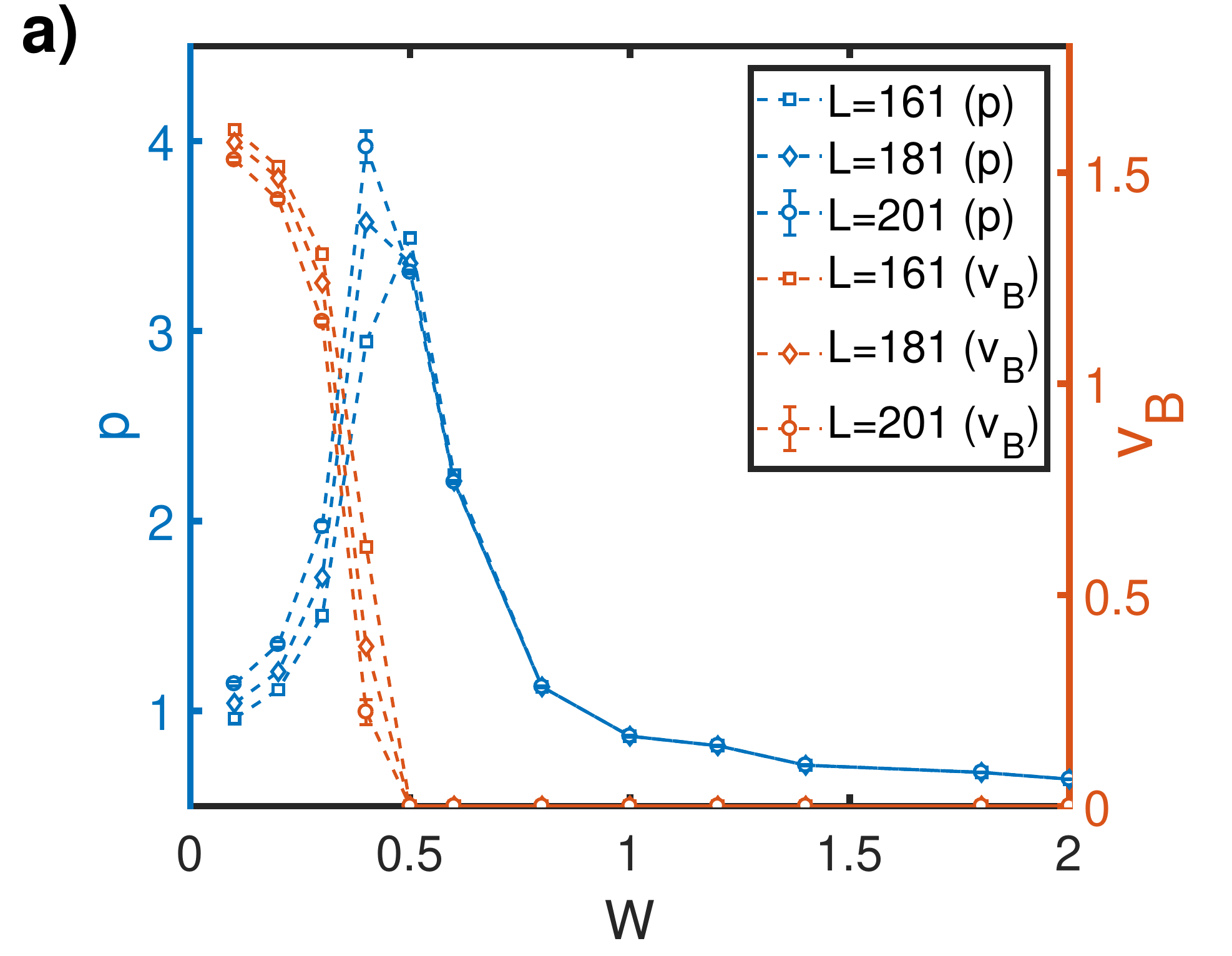}
		\includegraphics[width=0.7\columnwidth]{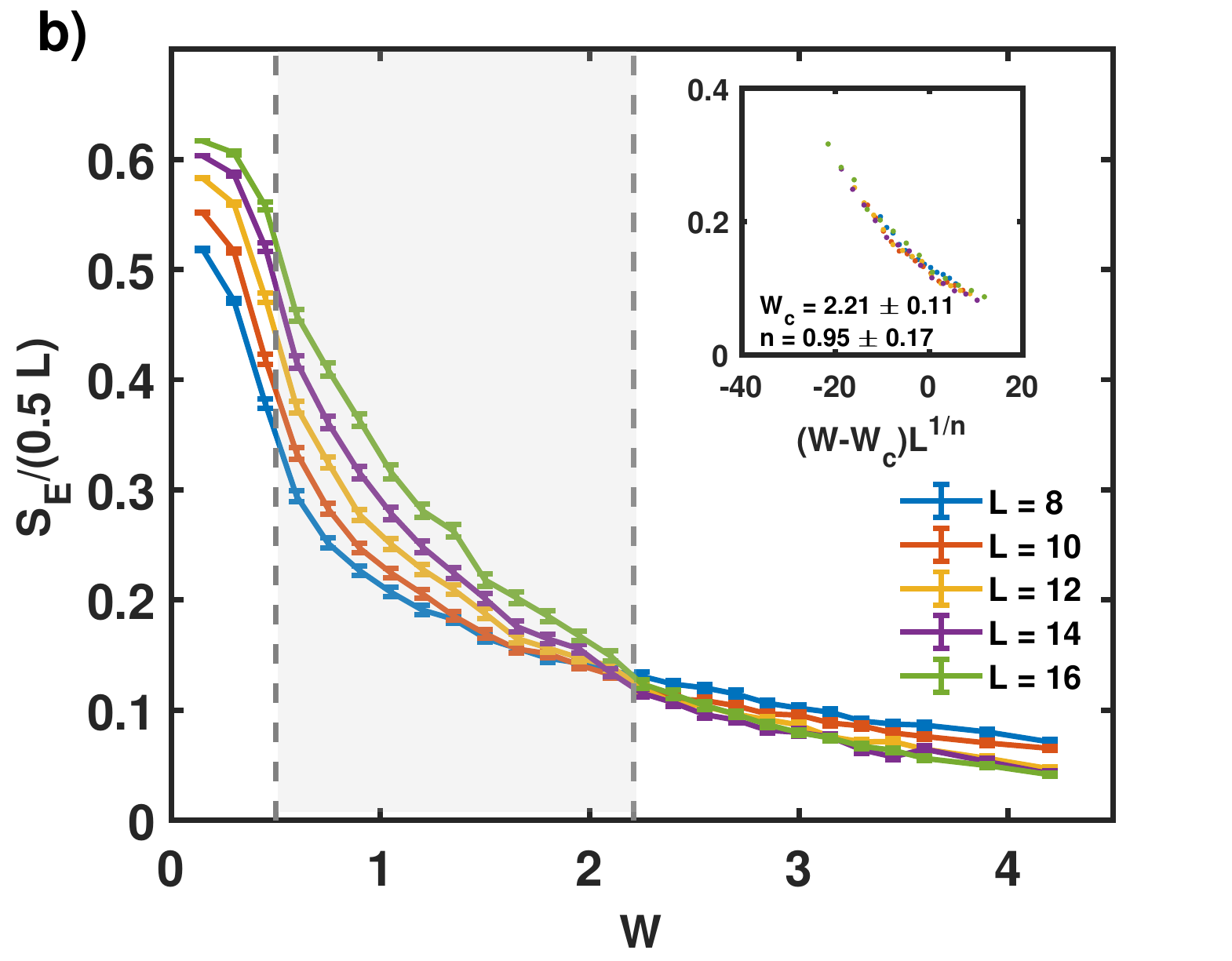}
		\caption{\textbf{a) }The extracted broadening coefficient $p$ and butterfly velocity $v_B$ are plotted for different sized systems, versus disorder. Note, $v_B$ goes to zero and $p$ has a peak at around disorder $W \sim 0.5$ with small finite-size effect. Errorbars obtained from the $95 \%$ confidence interval of fitting, are shown for the largest system size. \textbf{b) } Finite-size scaling on half-chain entanglement entropy estimates that the localization transition occurs at $W_c\sim 2.21$. The data collapse to the degree 3 polynomial ansatz $g[(W-W_c)L^{1/n}]$ with $n\sim 0.95$ is shown in the inset. The shaded region is the intermediate region.}
		\label{Fig:MFI_gaussian}
	\end{figure}
	
	We detect the transition by extracting the butterfly velocity and the wavefront broadening from the averaged squared commutator. We use the universal form for the squared commutator ahead of the wavefront (where $C(x,t)<<1$), conjectured in \cite{Xu2018, Khemani2018, Xu2018a}, 
	\begin{equation}\label{eq:cfitform}
	C(x,t)\sim \exp\left(-\lambda_p\left(x-v_B t\right)^{1+p}/t^p\right)
	\end{equation}
	Here, $v_B$ is the butterfly velocity, and $p$ is the wavefront broadening coefficient, which is known to be $p=1$ for random unitary circuit models \cite{Nahum2018,VonKeyserlingk2018}, $p=0$ for large-N holographic models and $p=\frac{1}{2}$ for non-interacting systems. The above form does not hold in the localized regime, which has a logarithmic lightcone \cite{Swingle2017a,Huang2017a,Fan2017,He2017,Chen2016,Chen2017,Slagle2017}. Additionally, the shape of lightcone becomes power-law like before the MBL transition due to rare region effects \cite{Luitz2017, Nahum2017}.  A general form that captures all the scenarios is,
	\begin{equation}\label{eq:cmegafitform}
	C(x,t)\sim \exp\left(-\lambda_p\left(x-v_B t\right)^{1+p}/t^p+a\log(t)\right)
	\end{equation}
	This form captures the cases where the lightcone is linear ($v_B \neq 0$, $a = 0$), power-law ($v_B =0$, $p \neq 0$, $a = 0$) or logarithmic ($p = 0$, $v_B = 0$, $a \neq 0$), as the disorder strength increases. 
	
	
	\textit{Numerical result -- } Here we use the mixed-field Ising model with Gaussian disorder as an example to  demonstrate the transitions in Fig.~\ref{Fig:phasediag}. The other three cases can be found in the Supplemental Material (S.M.).  In Fig.~\ref{Fig:MFI_gaussian}, we plot the extracted $v_B$ and $p$ versus disorder, for different lengths of the spin chain by fitting the data to the growth form \eqref{eq:cfitform}.  The fitting procedure and the goodness of fit are discussed in S.M., Sec. II.
	The butterfly velocity decreases as the disorder strength increases and becomes zero at $W\sim 0.5$. On the other hand, $p$ increases as $W$ approaches the critical disorder,  and decreases when $W$ passes beyond that. This disorder is below the MBL transition disorder extracted from exact diagonalization study on the entanglement entropy scaling (Fig.~\ref{Fig:MFI_gaussian}(b)). 
	The fact that $v_B$ goes to zero and $p$ peaks at the same disorder strength indicates a sharp transition before the true MBL transition, consistent with the weak-link model describing the rare region effects in disordered systems, studied recently~\cite{Nahum2017}.
	
	Below the transition, the system is characterized by a finite $v_B$ and $p$, indicating a linear lightcone with broadening front. Above the transition, the velocity becomes zero and the shape of the lightcone becomes powerlaw like, $x\sim t^ {p/(p+1)}$. Our method captures the logarithmic lightcone in the strong disorder limit (Fig.\ref{Fig:lightcone} (c)), but it is difficult to ascertain the transition to the logarithmic light cone from fitting the finite space-time data. This is discussed in S.M., Sec. II, where we also provide more evidence of logarithmic light cones at high disorder strength beyond the MBL transition. The transition identified here is different from the diffusive-subdiffusive transition for dynamics of conserved quantities \cite{Znidaric2016,Mendoza-Arenas2018}. In particular, we observe that in the Heisenberg model with box disorder, the $v_B=0$ transition occurs at a higher disorder, $W\sim4$ than the spin transport diffusive-subdiffusive transition disorder, $W\sim1.1$ (from \cite{Znidaric2016}, in our Pauli matrix convention). This implies a separation of information propagation and spin transport.

	\textit{Shot to shot variability -- }
	\begin{figure}
		\includegraphics[width=0.490\columnwidth]{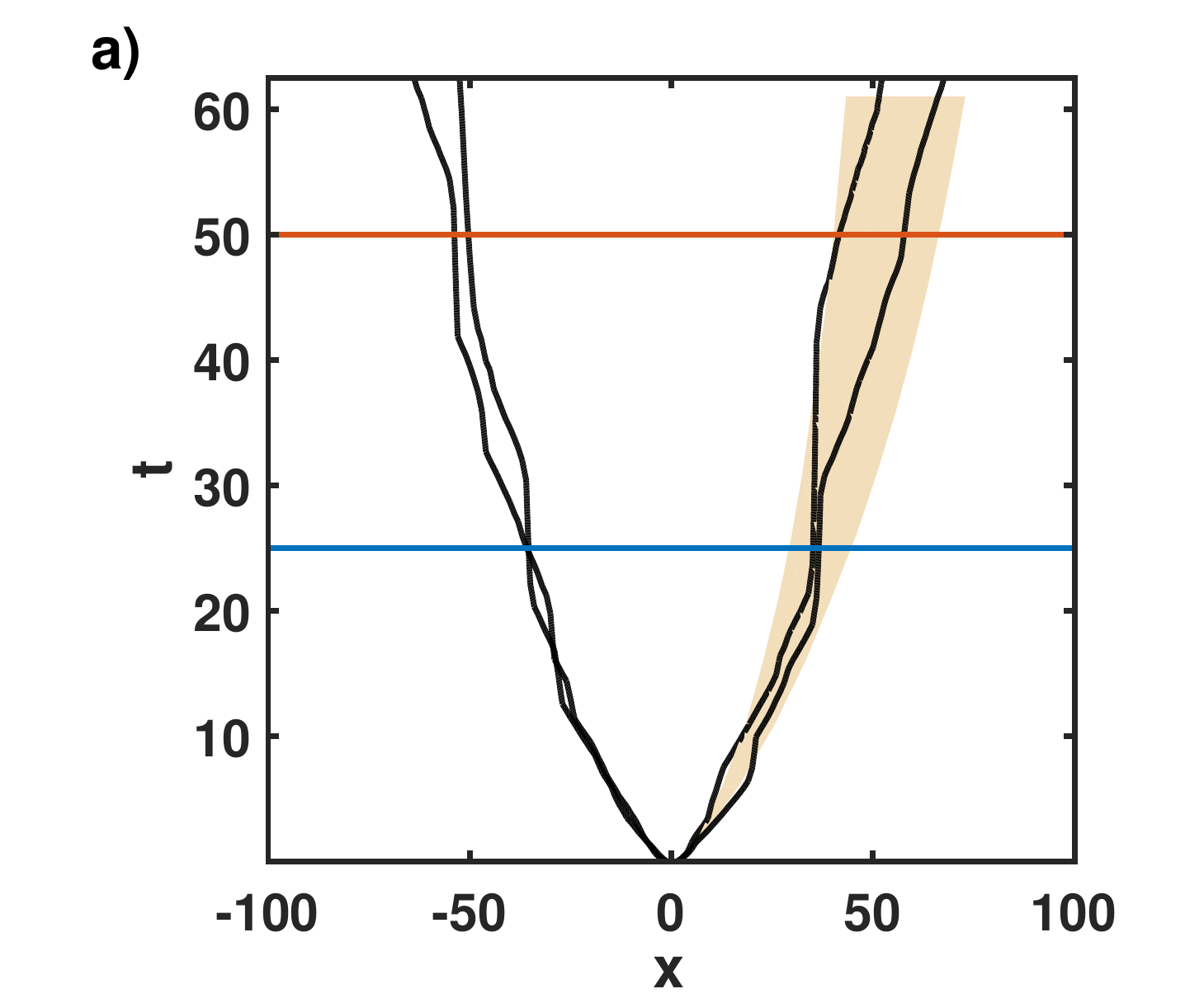}
		\includegraphics[width=0.475\columnwidth]{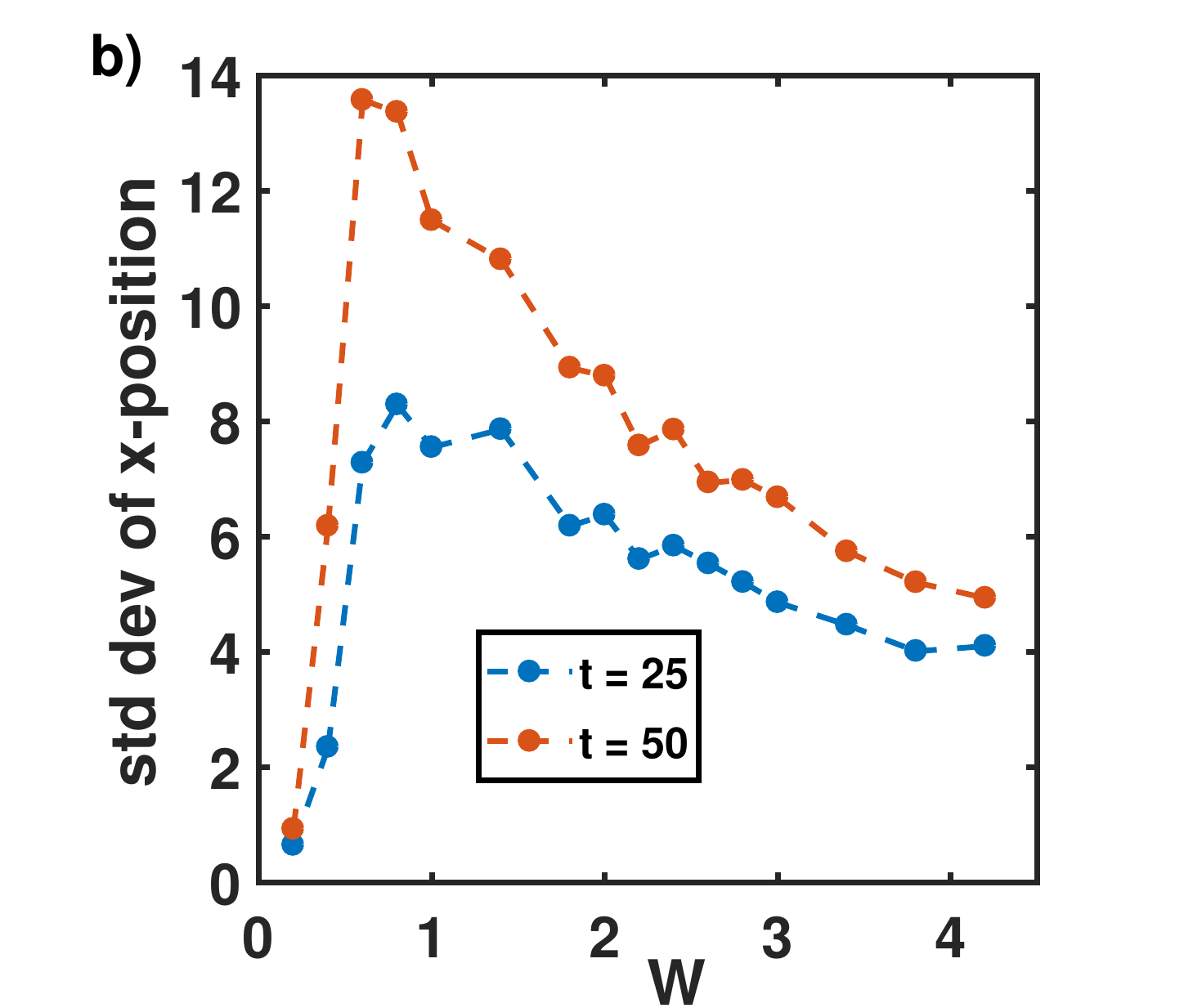}
		\caption{\textbf{a) }The bold black lines are \textit{single realizations} of $-15$ contour lines of $\log(C)$ at disorder $W=0.8$ for the mixed field Ising model with Gaussian disorder. Note the colored patch is given by the SD of the $x$ positions for 180 realizations at a given time. Note that the two disorder realizations have distinct behaviors after $t=25$, with one being significantly slower because of a local bottleneck of large disorder.  \textbf{b) }SD of $x$-cuts at times $t=25$ and $t=50$, for 180 realizations for different disorders are plotted, which peaks at $W\sim0.5$ and coincides with the critical disorder where $v_B$ vanishes.}
		\label{Fig:variance}
	\end{figure}
	We also study the variability of the contours of $\log(C)$ from one disorder realization to another. In Fig.~\ref{Fig:variance}(a) a particular contour line of $\log(C)$ is plotted for two different disorder realizations with $W = 0.8$， which differ significantly. To characterize the shot to shot fluctuations, in Fig.~\ref{Fig:variance}(b), we plot the SD of  $x$ positions, and observe that at long time, the variability peaks at the same disorder $(W\sim0.5)$ where $v_B$ vanishes.  The divergence of fluctuations,  obtained without any numerical fitting, is remarkably consistent with the divergence of $p$ in Fig.~\ref{Fig:MFI_gaussian}(a). This substantiates the transition at $W\sim0.5$. Fig.~\ref{Fig:variance}(a) also demonstrates the  microscopic mechanism for vanishing $v_B$ before the MBL eigenstate transition. The contours for two different realizations have bottlenecks at certain space regions, where scrambling is arrested. This is a visualization of rare region effects - local stronger disorders in certain regions affecting average dynamical properties. 
	
	 \textit{Rare region model -- } Motivated by above numerical results, we construct a simple model of rare regions which explains the emergence of power law, broadening behavior, and the existence of a ballistic phase at weak disorders. In a $L$ sized spin chain with Gaussian random disorders $\mathcal{N}(0,\sigma^2)$, the SD of local disorder, might be different from $\sigma$. It might also exceed the MBL critical disorder $\epsilon_{c}$, even when $\sigma<\epsilon_{c}$. Let $\epsilon$ be the disorder beyond which the operator growth has a logarithmic light cone. Consider a continuous stretch of $\alpha\log(L)$ spins, whose SD exceeds $\epsilon$. The balance between the exponentially slow transport and logarithmic size of such region leads to overall subballistic information transport. Specifically, the time it takes for the information to propagate across the chain with one such rare region is  $t \sim L/v_B+e^{\zeta\alpha \log L}$, where $\zeta$ is treated as the averaged inverse length scale associated with the logarithmic cone for the current purpose (It is defined carefully in S.M. Sec. V). In the limit $L\to\infty$, the average velocity $L/t$ goes to zero for $\zeta\alpha>1$, indicating the subballistic scenario. This corresponds to the case where the rare region is long enough that it dominates the time, $t \sim L^{\zeta \alpha}$. As the ballistic transition is approached, we have $\zeta\alpha\to1^{+}$. Comparing to the power-law lightcone $x\sim t^{p/(p+1)}$ indicates that $p\to\infty$, consistent with the apparent divergence of $p$ at the ballistic-subballistic transition in our numerical result. A related but distinct approach was considered in \cite{Nahum2017}, where the rare region effects on operator spreading were quantified using a coarse grained quantity related to the entanglement spreading across weak-links. Our model is directly in terms of the bare disorder and gives rise to consistent predictions. 
	 
	 The existence of a ballistic phase in the low disorder limit is also borne out of the simple model. Consider the probability of having \textit{no} rare region of length $\alpha \log L$ with SD larger than $\epsilon$ in a disordered spin chain of length $L$ with global SD $\sigma$, denoted as $q(\alpha;\sigma,\epsilon)$. In general, $q$ decreases with $\sigma$ and increases with $\alpha$.  Based on the above discussion, any $\alpha$ larger than $1/\zeta$ leads to subballistic slowing down of the information propagation. Therefore, a sufficient condition for ballistic propagation is that no such disruptive rare regions occur, i.e., $q(1/\zeta;\sigma,\epsilon)=1$. In  Sec. V of S. M., we prove the following inequality,
\begin{equation}
	q(1/\zeta;\sigma,\epsilon) \geq  \left(1-\beta^{\log(L)/\zeta}\right)^{{\frac{\zeta L}{\log(L)}}}
	\label{eq:badprob2}
\end{equation}
	where $\beta = \left(\frac{\epsilon^2}{\sigma^2}e^{1-\frac{\epsilon^2}{\sigma^2}}\right)^{1/2}$. In the limit, $L\to \infty$, the RHS of Eq.~\ref{eq:badprob2} is $1$ when $\beta<e^{-\zeta }$. In terms of microscopic parameters, the condition becomes,
	\begin{equation}\label{eq:ballisticcondition2}
	\frac{\epsilon^2}{\sigma^2}e^{1-\frac{\epsilon^2}{\sigma^2}}<e^{-2\zeta}
	\end{equation}
	Since $\zeta$ is finite, there exists a finite $\sigma^*$, below which  all $\sigma$ satisfy the sufficient condition for ballistic transport Eq.~\ref{eq:ballisticcondition2},  leading to a finite window of a ballistic phase.
	
	It is worth noting that the model only shows the existence of a ballistic phase for $\sigma<\sigma^{*}$. The inequality is a sufficient, but not a necessary condition for ballistic transport; hence $\sigma^{*}$ should not be mistaken with the critical ballistic-subballistic transition. Furthermore, in our numerics, we can't resolve $\epsilon$, where sub-ballistic becomes logarithmic (in a finite system data, a soft power law is difficult to resolve from a logarithm), or $\zeta$ which will be a complicated averaged scale. Hence we can't quantitatively verify Eq.~\ref{eq:ballisticcondition2}. A more careful study of the difference between the average time $\overline{t}$ and the typical time $\exp(\overline{\log t})$ should be considered to further characterize the ballistic to sub-ballistic transition.

	\textit{Conclusions -- }
	We studied the ballistic to sub-ballistic crossover in operator spreading for large interacting disordered spin systems using MPO dynamics, for different spin Hamiltonians and error models. Our numerical results establish the existence of a ballistic phase and a sharp transition to a subballistic phase. The numerical observation of fluctuations of the wavefront motivate a simple model of rare regions which explains aspects of this transition. Natural extensions of the rare region model would be to incorporate the effects of wavefront broadening into the analysis. Also our work demonstrates a separation between information propagation and spin transport \cite{Znidaric2016,Mendoza-Arenas2018}, which could be an interesting direction of future study.

\textit{Acknowledgements:} We acknowledge University of Maryland supercomputing resources (http://hpcc.umd.edu), specifically the Deepthought2 cluster, used in this work. We also thank D. Huse, V. Khemani, and S. Gopalakrishnan for discussions. This material is based on work supported by the Simons Foundation via the It From Qubit Collaboration, by the Air Force Office of Scientific Research (FA9550-17-1-0180), and by the NSF Physics Frontier Center at the Joint Quantum Institute (PHY-1430094).
	
	\begin{appendices}
		\section{Convergence with bond dimension}
		\label{app:convergence}
		In this appendix we demonstrate convergence with bond dimension for the squared commutator data. In Ref.~\cite{Xu2018}, it was rigorously proven that if $C(x,t)$ is sufficiently small for all $x > x_0$, then the operator Renyi entropy with entanglement cut at $x_0$ is also small. This result implies that the MPO representation with a fixed finite bond dimension is faithful for operators of physical importance. There is still a possibility that errors could build up after repeated truncations, but it was also argued that these errors cannot propagate outside the emergent light cone.
		\begin{figure}	
			\includegraphics[width=\columnwidth]{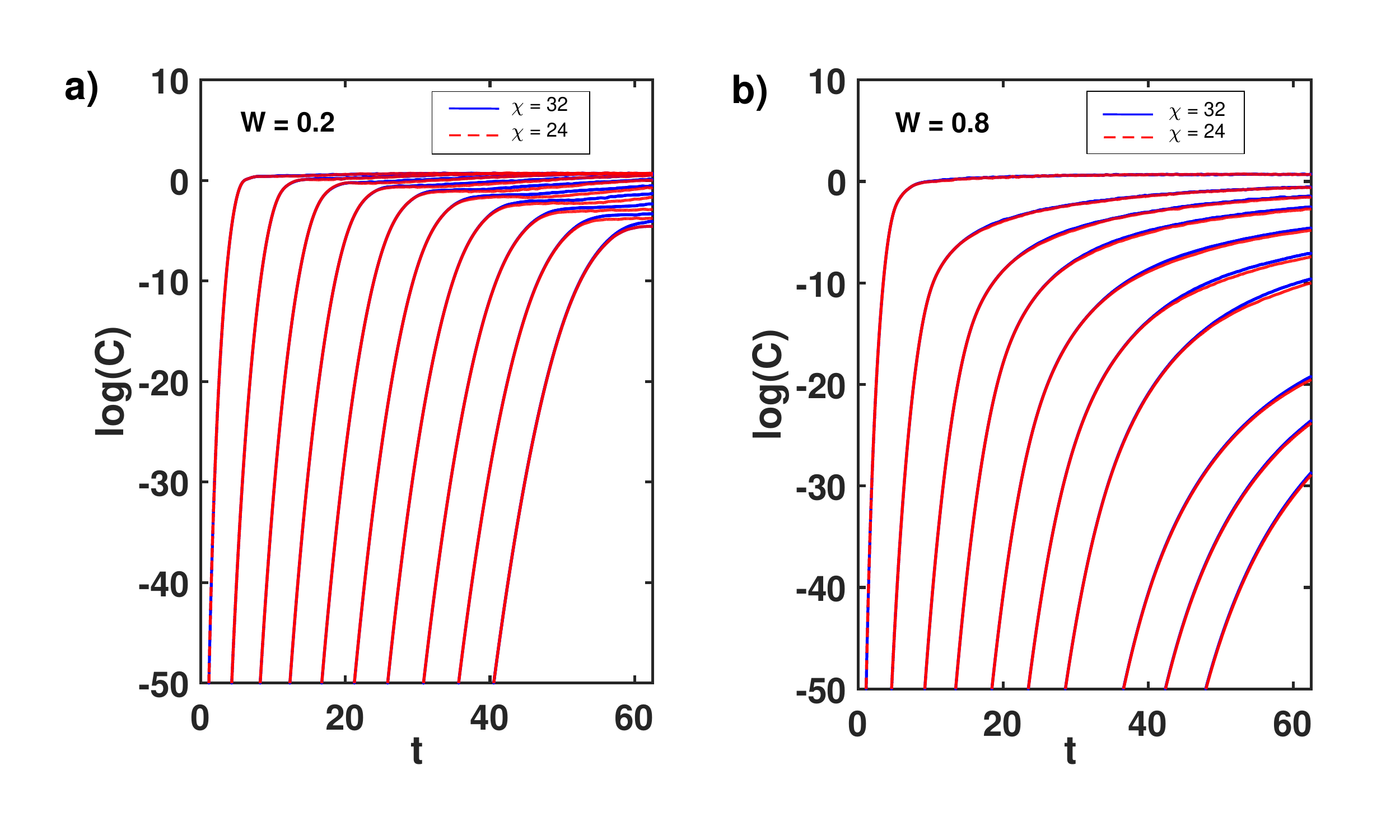}		
			\caption{The logarithm of the squared commutator, $\log(C)$, for single realization with \textbf{a)} $W=0.2$ and \textbf{a)} $W=0.8$. The plots are for two bond dimensions, $\chi=32$ (continuous blue line), and $\chi=24$ (dotted red line). They are essentially indistinguishable.}
			\label{Fig:Conv1}
		\end{figure}
		\begin{figure}
			\includegraphics[width=\columnwidth]{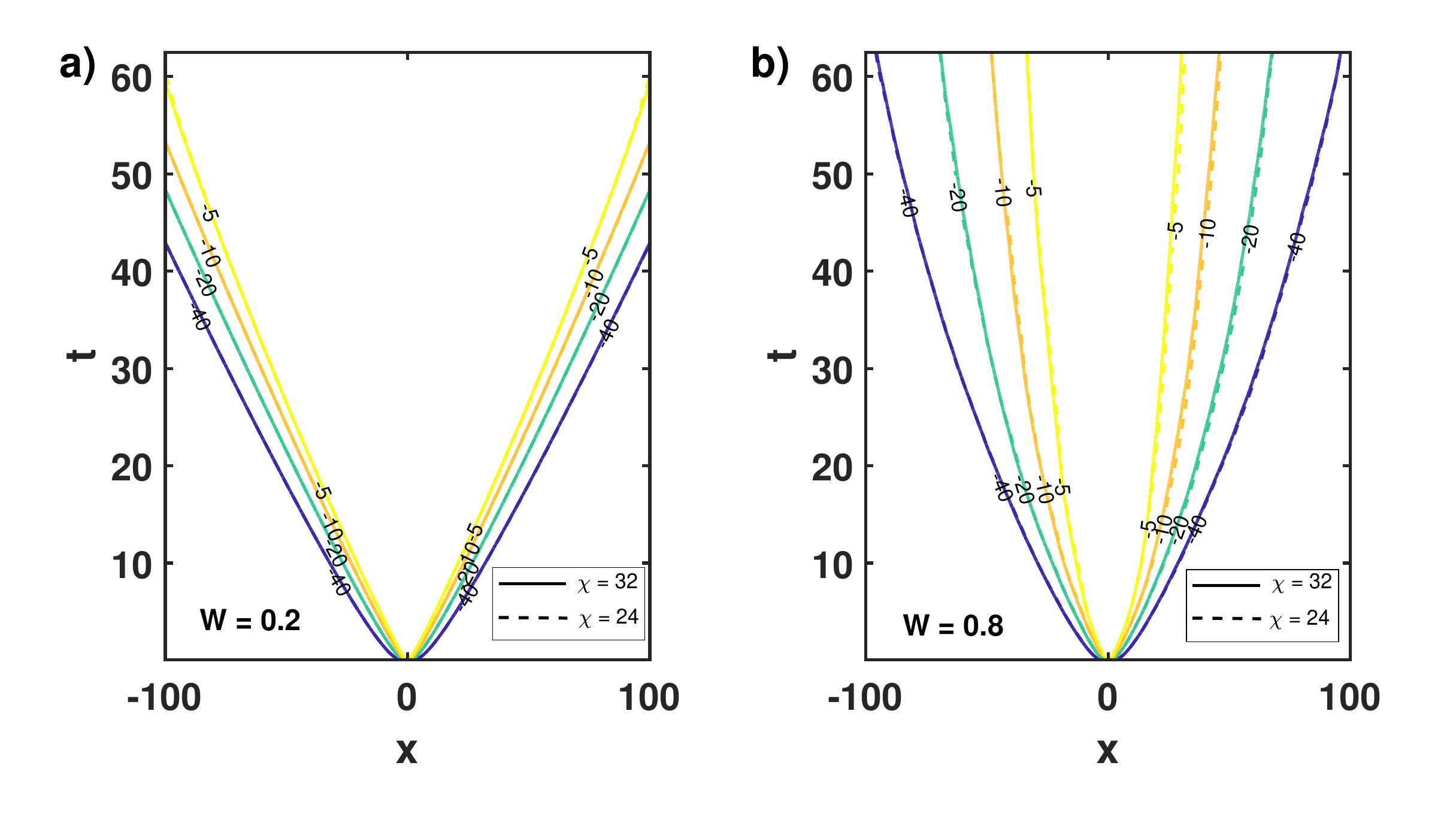}
			\caption{The contours of averaged $\log(C)$ (averaged over $500$ disorder realizations) for two different disorders \textbf{a)} $W=0.2$ and \textbf{b)} $W=0.8$ for the mixed field Ising model with Gaussian disorder are plotted. The continuous lines are for $\chi=32$, while the dotted lines are for $\chi=24$.}
			\label{Fig:Conv2}
		\end{figure}
		
		In a many-body localized system, the light cone grows logarithmically instead of linearly with time, and thus one hopes to access an even wider region of the space-time with this method. In that sense, MBL is easier than chaos, as the spatial spread is less. In the chaotic case, the linear light cone ensures that errors within the light cone are contained within, but in the logarithmic case, the error containment is not so straight forward. Due to these two opposing factors, we need to numerically study the convergence of the light-cones with increasing bond dimension. We consider an $L=201$ spin chain, and look at the overlap of $X_{r=101}(t)$ with $X_r$ as a function of $t$. In
		Figs \ref{Fig:Conv1} and \ref{Fig:Conv2}, we show convergence of both the single realization and the averaged data of $\log(C)$ with increasing bond dimension ($\chi=24$ and $\chi=32$) respectively. The data shown here corresponds to the mixed field Ising model with Gaussian disorder, which was considered in the main letter. Since the obtained data converges well (for system sizes and times considered) the rest of the numerical results shown in this paper have been obtained from MPOs with bond dimension $\chi=32$.
		\section{Extracting butterfly velocity and the logarithmic lightcone}
		\label{app:fitting}
		\begin{figure}	
			\centering
			\includegraphics[width=\columnwidth]{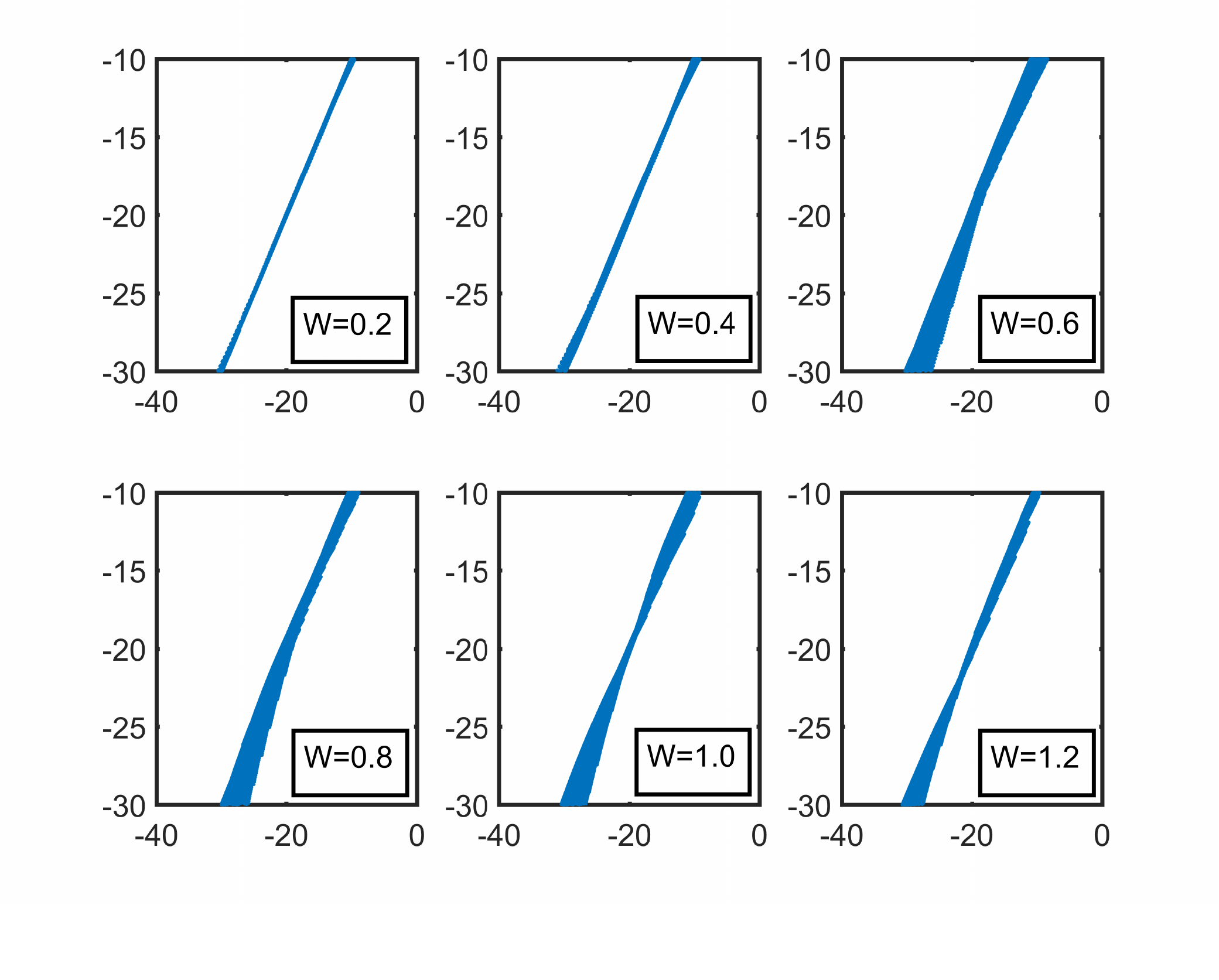}
			\caption{The averaged $\log(C)$ is plotted against the fitting ansatz $(x-v_B t-x_0)^{1+p}/t^{p}+c/a\log(t)$ for different disorders $W=0.2,0.4,...,1.2$, for the mixed field Ising model with Gaussian disorder. The fitted parameters for the figure are given in Table~\ref{Table:fittedparameter}.}
			\label{Fig:goodnessfit}
		\end{figure}
		
		To extract the physically relevant quantities from our numerical data, we employ a fitting procedure, in which we fit the disorder averaged $\log(C(x,t))$ to the fitting ansatz, $$\log(C(x,t))\sim a (x-v_B t-x_0)^{1+p}/t^{p}+c\log(t)$$ Note, the free parameters are $a$,$c$, the offset $x_0$, butterfly velocity $v_B$ and the wavefront broadening coefficient $p$. We fit the averaged data over a large domain, $-30<\log(C)<-10$, for which we are certain that the numerical procedure converges, to this ansatz, with the physical constraints $v_B\geq0$ and $p\geq0$. The collapse of the data to this fitting form is demonstrated in Fig.~\ref{Fig:goodnessfit}. The fitted parameters for the figure are given in Table~\ref{Table:fittedparameter}.
		\begin{table}
			\begin{tabular}{|c|c|c|c|c|c|}
				\hline
				W&$v_B$ &$p$ &$a$ &$c$&$x_{0}$ \\ \hline
				0.2&1.4357&1.3501&-1.1903&0.0452&0.1075\\ \hline
				0.4&0.2227&3.9690&-0.0278&0.0000&-0.1005\\ \hline
				0.6&0.0000&2.2058&-0.0867&0.0000&-0.1461\\ \hline                  
				0.8&0.0000&1.1273&-0.2530&0.0000&0.3807\\\hline
				1.0&0.0000&0.8676&-0.3445&0.0000&1.8693\\\hline
				1.2&0.0000&0.8174&-0.3882&0.0000&1.8854\\                              
				\hline
			\end{tabular}
			\caption{Fitted parameters for Fig.~\ref{Fig:goodnessfit}}
			\label{Table:fittedparameter}
		\end{table}

		The fitting ansatz that we employ has the merit of capturing various possible scenarios of operator growth. From the chaotic growth considered in \cite{Xu2018} and \cite{Xu2018a} we expect $v_B>0$ and some finite $p$ for the situation without disorder. In the presence of weak disorder, there could be multiple possible options, one is that any weak disorder is enough to take $v_B$ to zero (as was indicated in \cite{Luitz2017}), or, there could be a phase in the ergodic side which could have $v_B>0$, as was argued in \cite{Nahum2017}. Furthermore, the behavior of the wavefront broadening in the presence of disorder is also not well understood. From the result of our numerical fitting procedure, we definitely see evidence of a ballistic phase in the presence of weak disorder, and furthermore, in the ergodic phase preceding the MBL transition, we observe a sharp transition at which $v_B$ goes to zero and the broadening coefficient $p$ seemingly diverges. The result doesn't change even if we remove the $\log$ term from the fitting ansatz, as its coefficient in the ergodic side has been observed to be vanishingly small.
		
		The fitting ansatz could also potentially capture the logarithmic lightcone in the MBL side. One possible way in which that can be achieved in the fitting ansatz is where $v_B=0$, $p=0$ and the coefficient of the $\log$ term is non zero. However we don't observe a sharp transition for the domain of disorders that we consider, possibly because the transition of a soft power law to logarithm is a invisible to the numerical fitting procedure given the finite domain.
		\begin{figure}	
			\centering
			\includegraphics[width=0.8\columnwidth]{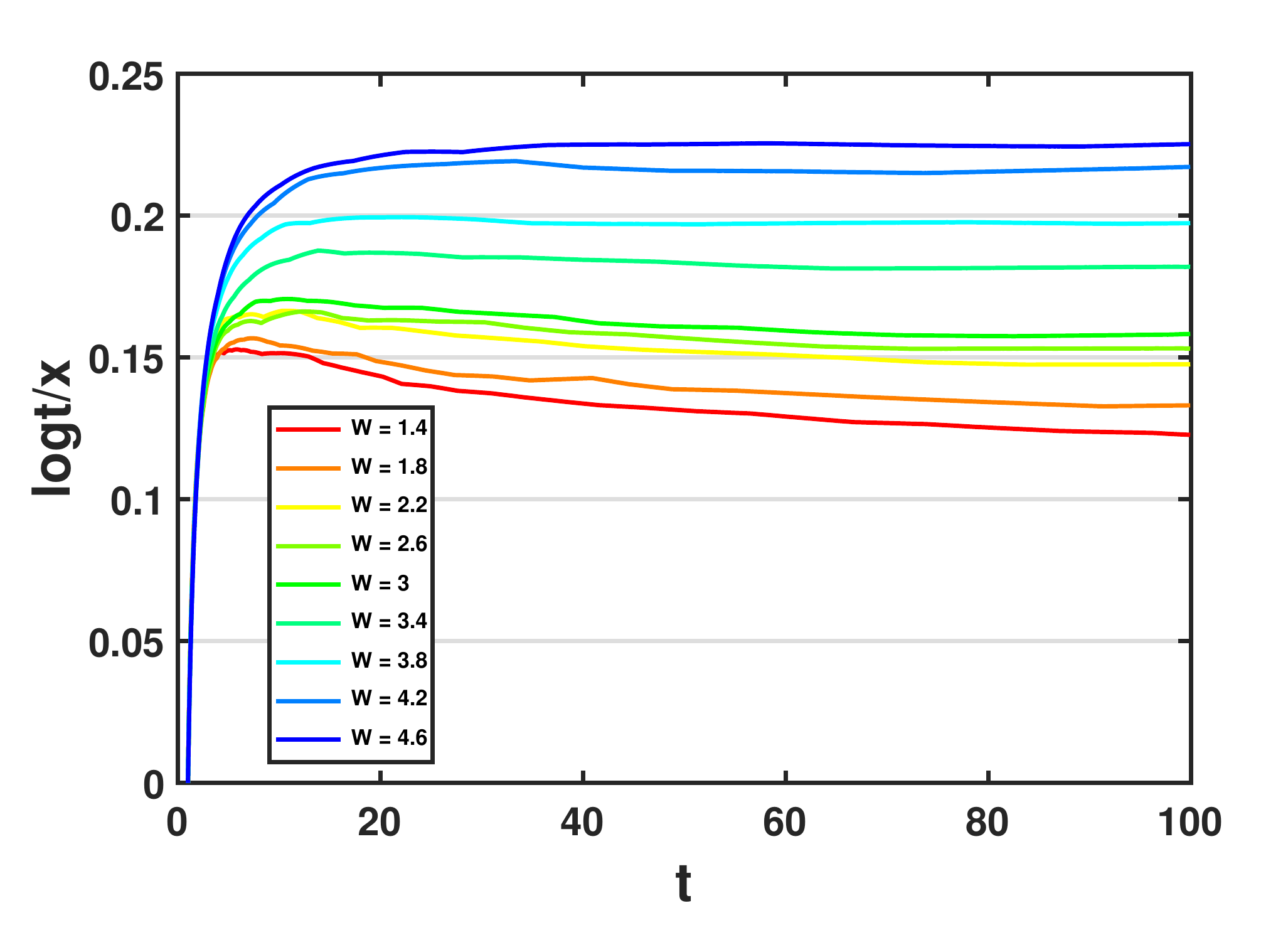}
			\caption{The $-10$ contour of the averaged $\log(C)$ for $201$ sized chain is extracted at different disorders, and $t$ and $x$ coordinates are obtained. $\log(t)/x$ is plotted against t, for different disorders for the mixed field Ising model with Gaussian disorder. At strong disorders, $W\gtrsim3.4$, the asymptotically flat plots provide evidence of a logarithmic light cone.}
			\label{Fig:logarithmiclightcone}
		\end{figure}
		
		In Fig.~\ref{Fig:logarithmiclightcone}, we show evidence of the logarithmic lightcone without using any numerical fitting procedure. We consider a particular contour ($-10$ contour of  $\log C$), and extract its $x$ and $t$ coordinates, and plot $\log(t)/x$ versus $t$ for different disorders. If the contour is logarithmic, the plot should approach a fixed value monotonically from below, and shouldn't decrease at late times. On the other hand, if the contour has a power law behavior, the plot will decrease with time. In Fig \ref{Fig:logarithmiclightcone}, we indeed see that for high disorders ($W \gtrapprox 3.4$) the plot is aympotically flat (note the long times considered, $t=100$). This provides evidence that at those disorders, the light cone is indeed logarithmic.
		
		\section{Heisenberg model and relation to diffusion}
		\label{app:HMD}
\begin{figure}	
			\centering
			\includegraphics[width=0.50\columnwidth]{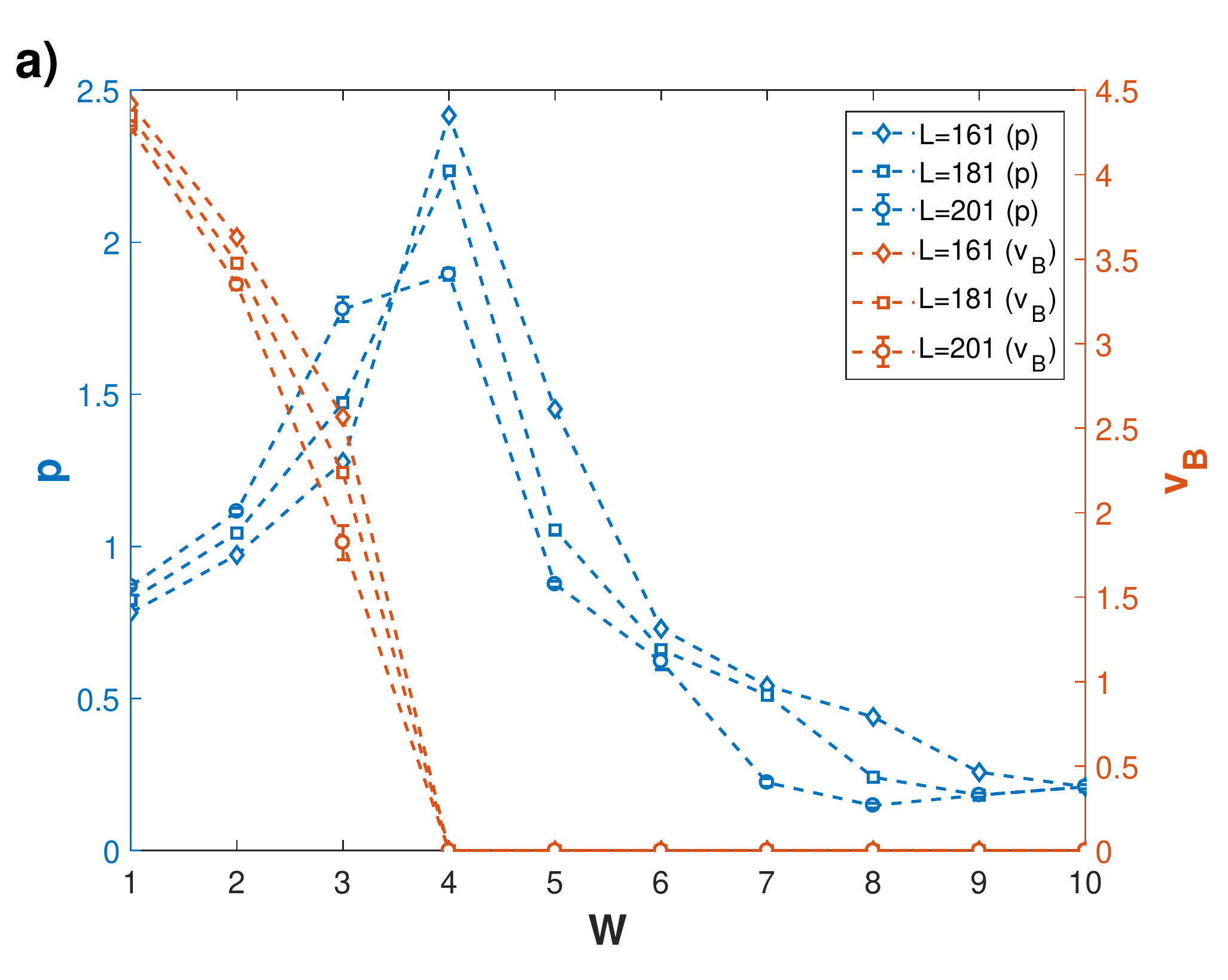}
			\includegraphics[width=0.45\columnwidth]{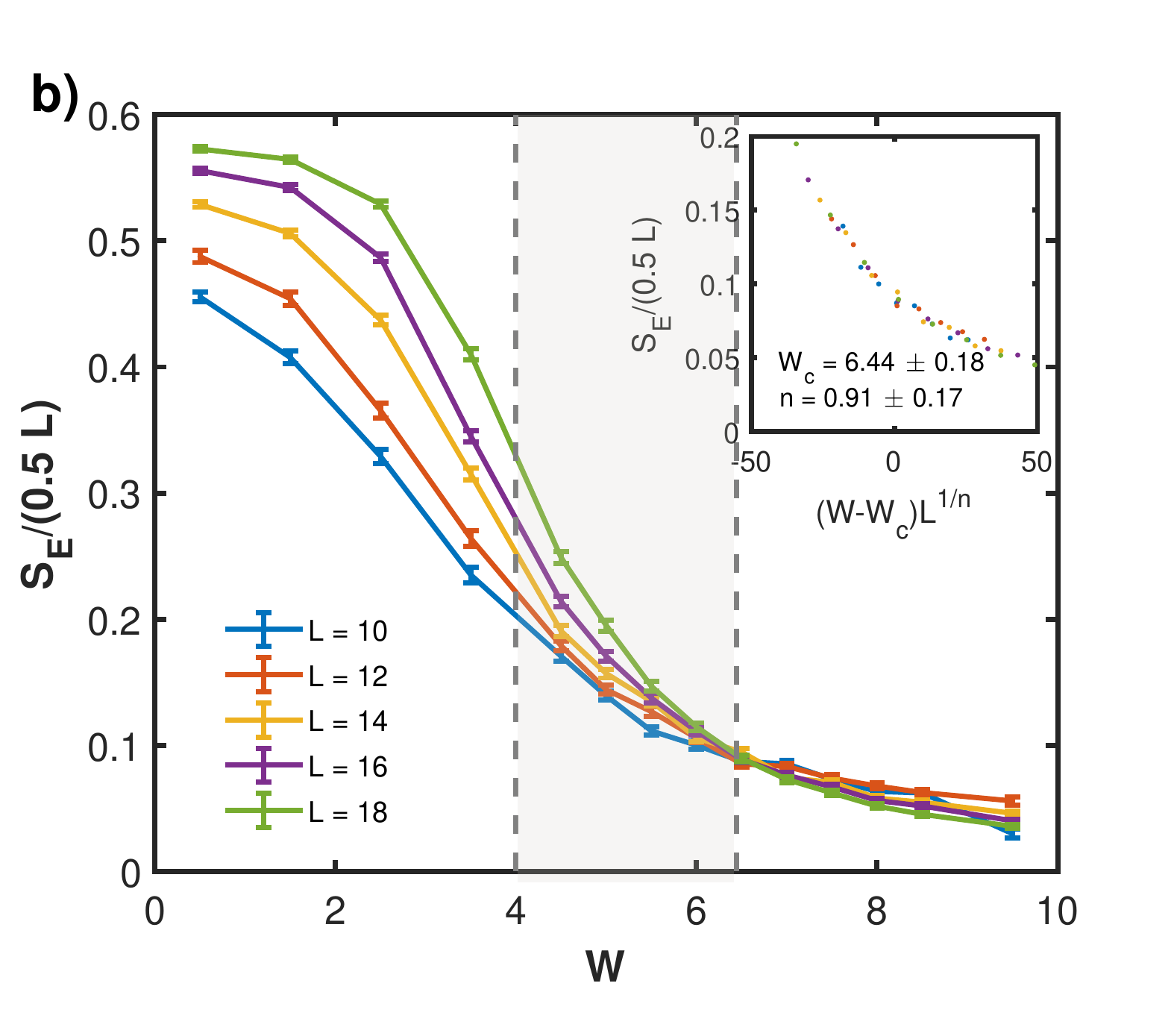}
			\caption{\textbf{a)} The extracted butterfly velocity $v_B$ and broadening coefficient $p$ are plotted for different sized systems of the disordered Heisenberg model (box), versus disorder. Note, $v_B$ goes to zero and $p$ has a peak at around disorder $W \sim 4$. Errorbars corresponding to the $95 \%$ confidence interval of fitting are shown for the largest system size.\textbf{b)} The MBL transition disorder is shown to be $W_c\gtrsim6.44$, which implies the shaded intermediate region which has powerlaw lightcones.}
			\label{Fig:HMD_box}
		\end{figure}

		We consider the Heisenberg model with box disorder. The fitted $v_B$ and $p$ are shown in Fig \ref{Fig:HMD_box}. This also shows $v_B$ going to zero and $p$ diverging at a disorder $W\sim 4$, which is lower than the MBL transition disorder, which has been extensively studied, and is known to be $\gtrsim 7$ \cite{Luitz2015b}.
		\par
		A related but distinct question is to study the dynamics of conserved quantities in the thermal regime in the presence of disorder. In \cite{Znidaric2016}, a transition between diffusive and subdiffusive transport was observed numerically in the Heisenberg chain, in the thermal phase. Corrected for the conventions used in the Hamiltonian we are considering, that transition occurs at $W\approx1.1$, which is not where we get the $v_B$ to go to zero. So this observation implies that there are two distinct transitions in the thermal side of the disordered phases, one for diffusive to sub-diffusive transport (which happens at smaller disorder), and the other between the ballistic and sub-ballistic operator spreading.
		
		\section{Comparison between box and Gaussian disorder}
		\label{app:boxgaussian}
		In this section we show the results of our analysis for the other two disorder models that we considered, which complete the phase diagram in our main paper.
		\begin{figure}	
			\centering
			\includegraphics[width=0.50\columnwidth]{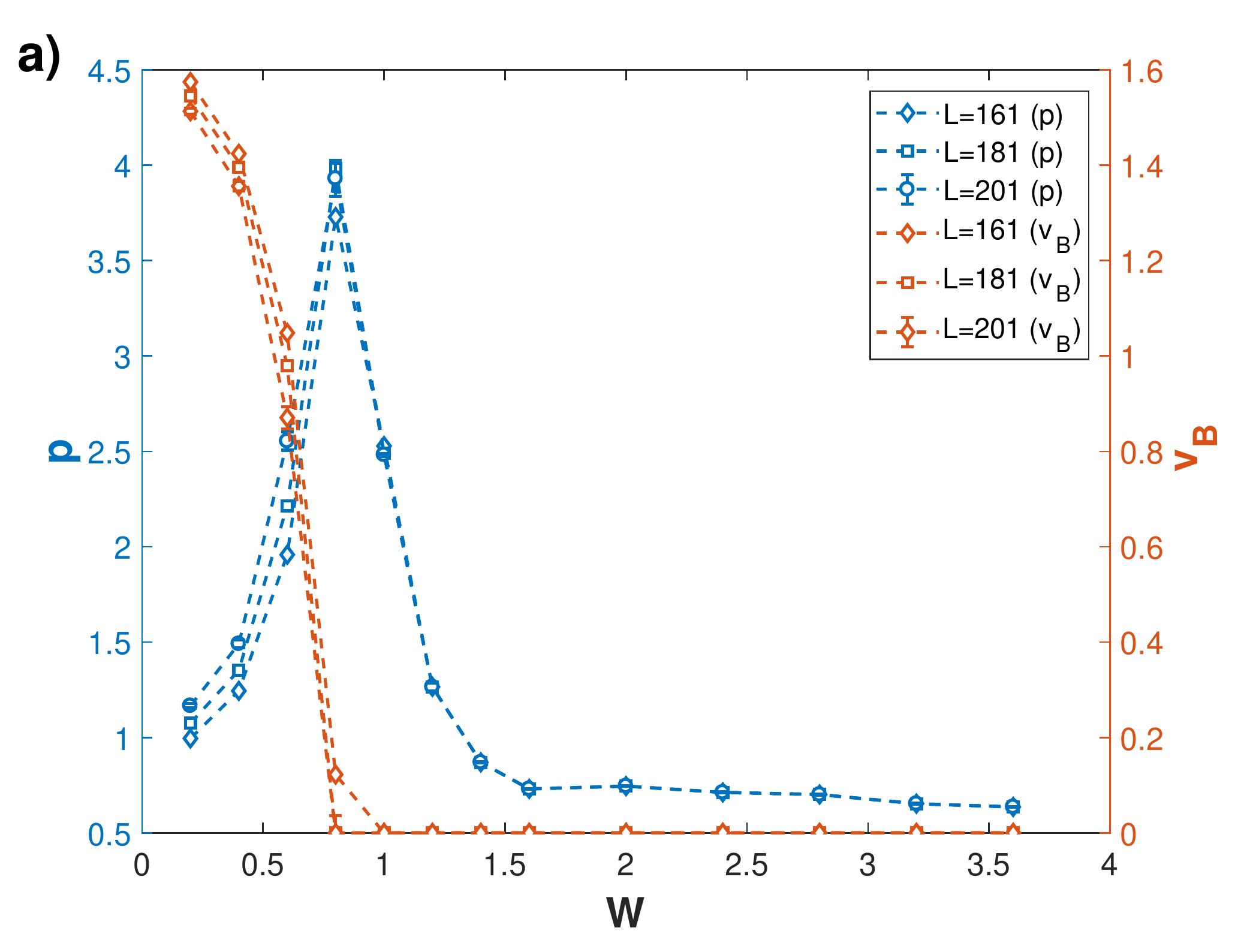}
			\includegraphics[width=0.45\columnwidth]{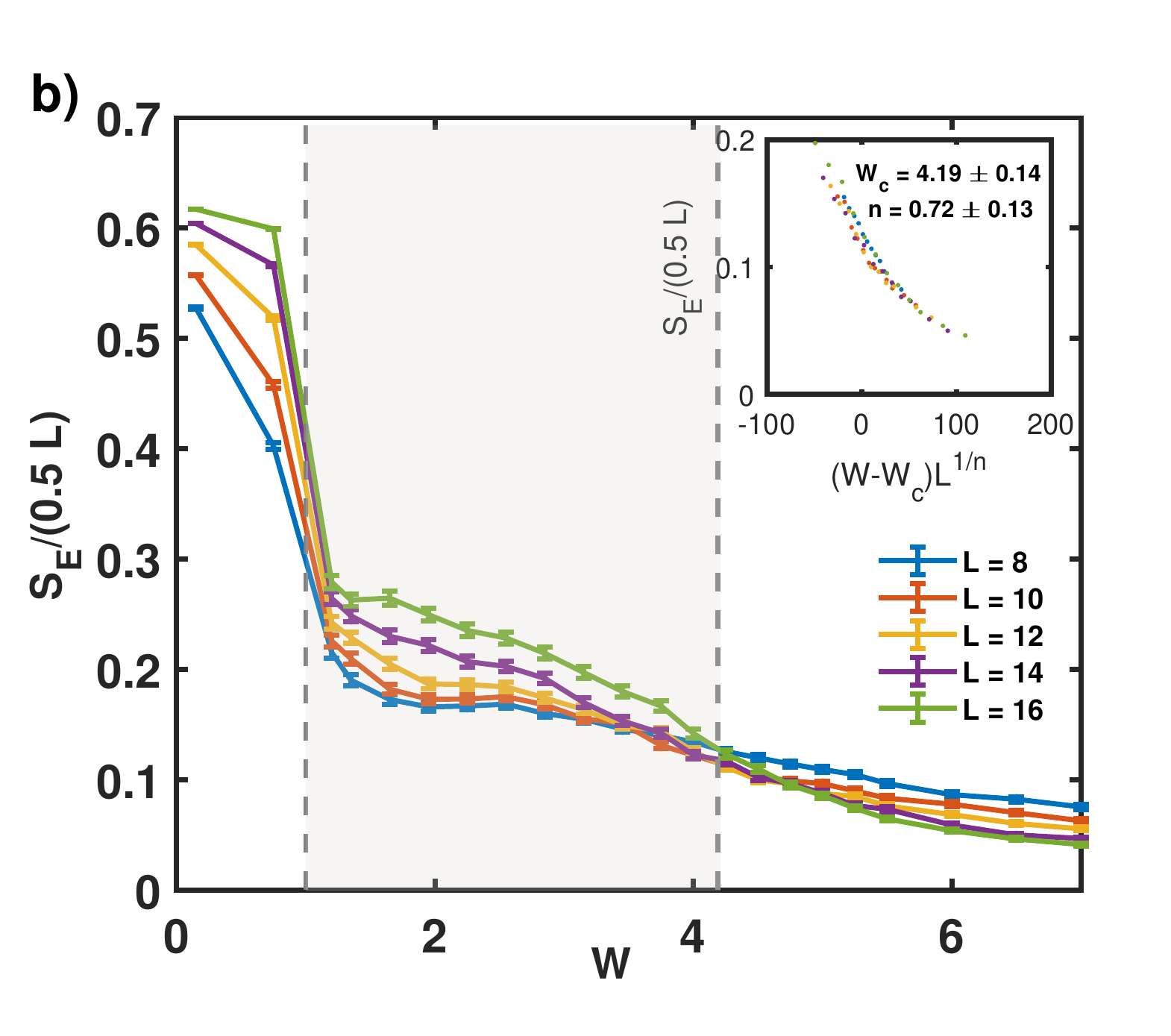}
			\caption{Results for the Mixed Field Ising model with box disorder. \textbf{a)} Extracted Butterfly velocity and broadening coefficient. Errorbars corresponding to the $95 \%$ confidence interval of fitting are shown for the largest system size. \textbf{b)} MBL transition ascertained by small system exact diagonalization.}
			\label{Fig:MFI_box}
		\end{figure}
		\begin{figure}	
			\centering
			\includegraphics[width=0.50\columnwidth]{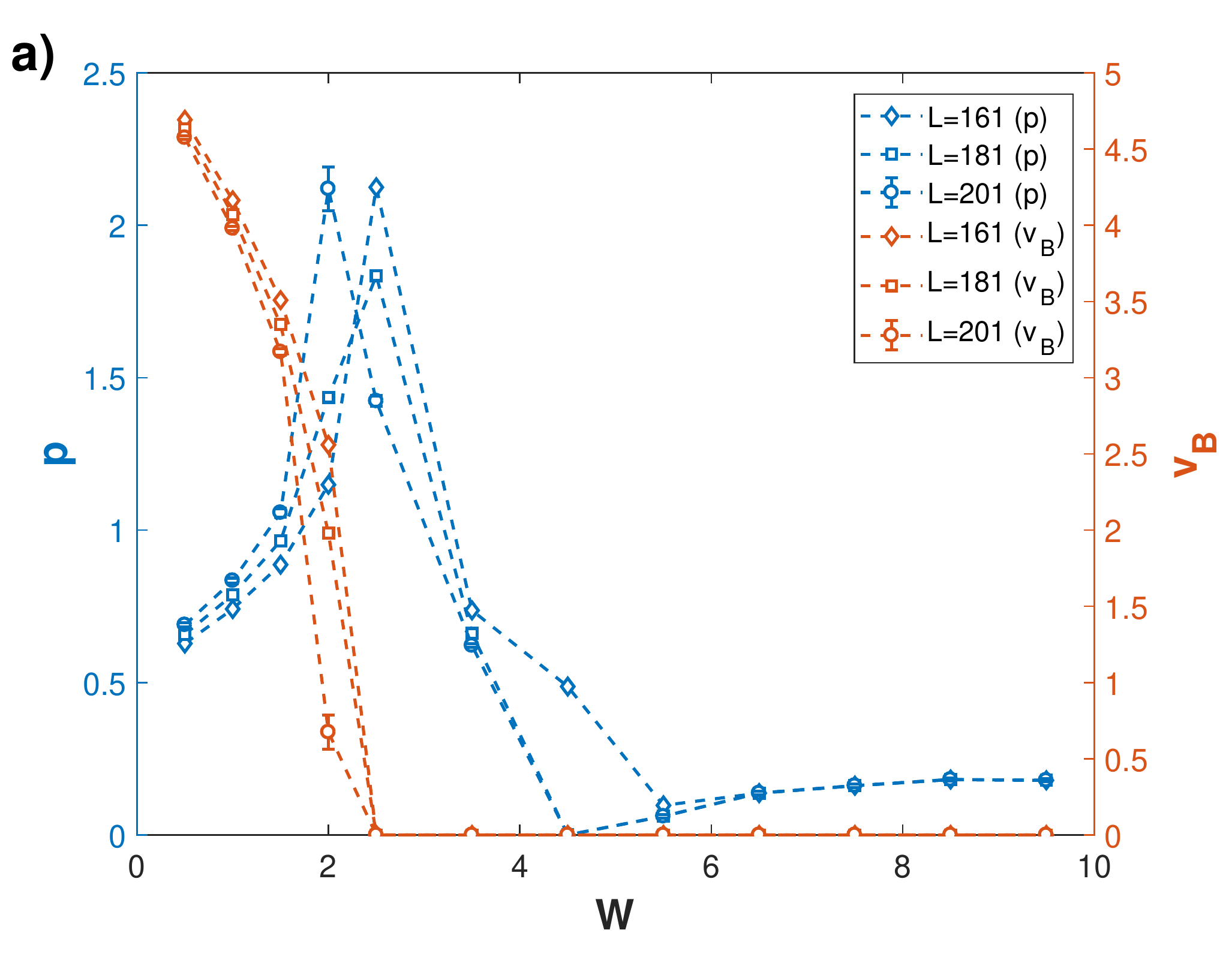}
			\includegraphics[width=0.45\columnwidth]{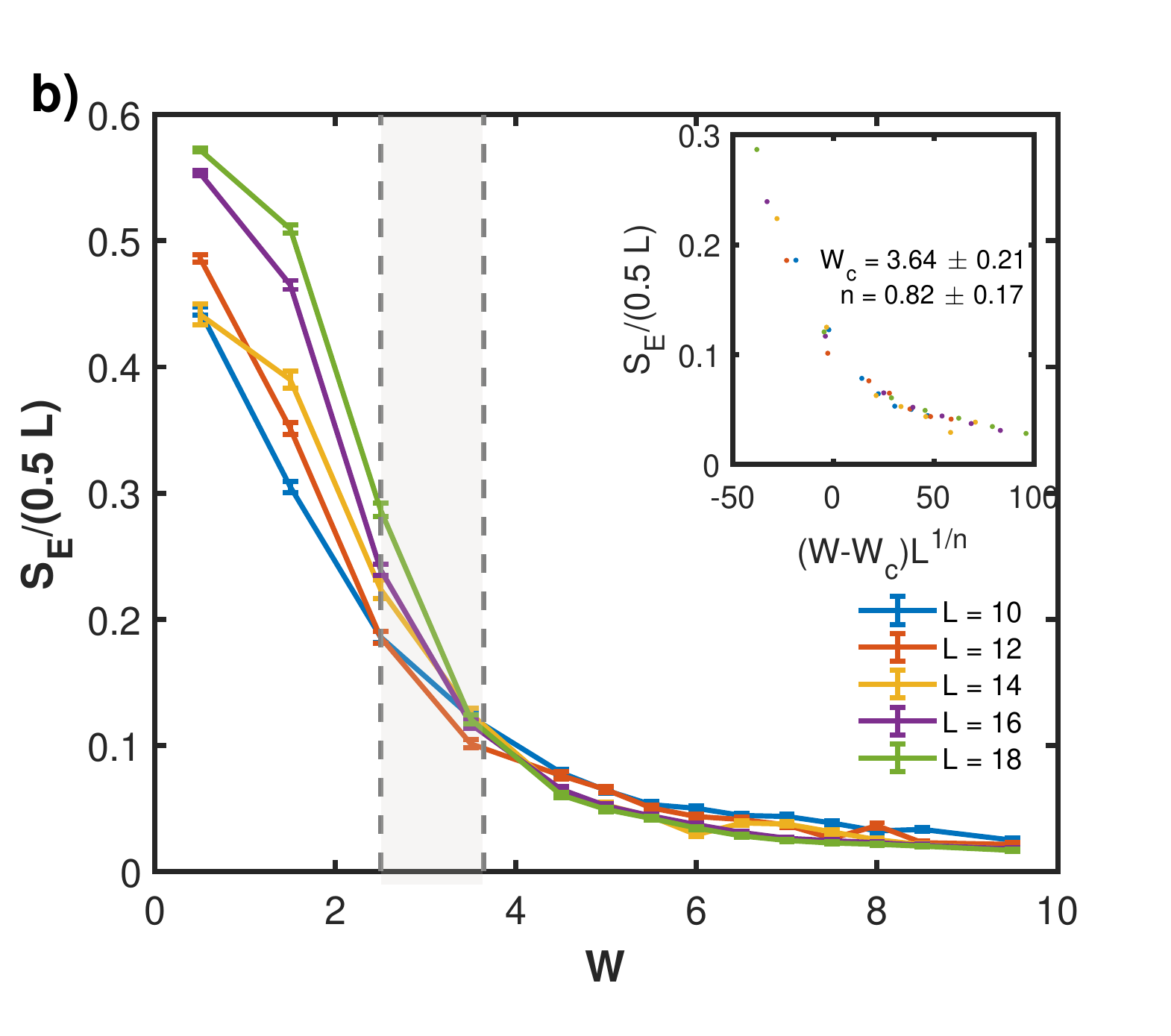}
			\caption{Results for the Heisenberg model with Gaussian disorder. \textbf{a)} Extracted Butterfly velocity and broadening coefficient.Errorbars corresponding to the $95 \%$ confidence interval of fitting are shown for the largest system size. \textbf{b)} MBL transition ascertained by small system exact diagonalization.}
			\label{Fig:HMD_gaussian}
		\end{figure}
		
		These results (Figs. \ref{Fig:MFI_box}, \ref{Fig:HMD_gaussian}) demonstrate the versatility of the numerical procedure employed, and also indicates how rare regions affect the thermalization-localization transition. Gaussian disorders allow for rare fluctuations more occasionally than box disorders, which results in onset of sub-ballistic transport and localization at lower disorders for the Gaussian case than the case with box disorder.
		
		\section{Calculations for the rare region model}
		\label{app:badbubble}
		In this section we prove Eq.7 in our paper, and explain the calculations for the rare region model we considered here. After proving Eq.7, we explain the meaning of $\zeta$, the averaged inverse length scale associated with the logarithmic light cone.
		
		The sample variance of normal random variables $\mathcal{N}(0,\sigma^2)$ satisfy the Chi-squared distribution. Hence for $n$ normal random samples we have
		\begin{equation}
		(n-1)\frac{s_n^2}{\sigma^2}\sim \chi^2_{n-1}
		\end{equation}
		where $\chi^2_{n-1}$ is the Chi-squared distribution of $(n-1)$-th order and the sample variance is defined as,
		\begin{equation}
		s_n^2=\sum\limits_{i=1}^{n}\frac{(x_i-\bar{x})^2}{n-1}
		\end{equation}
		
		We are interested in finding the probability that a $n$-sized region has locally larger variance than the putative critical disorder. Hence, the probability of a sample of size $n$ (in our picture, a continuous region of $n$ spins) having variance exceeding the subballistic-logarithmic critical transition strength $\epsilon^2$ is obtained from the cumulative distribution function of the Chi-squared distribution,
		\bea
		p(n;\sigma,\epsilon) & = \text{Prob}(\sigma_n^2\geq\epsilon^2)\\
		& = 1-\chi^2_{(n-1)|CDF}\left(\frac{n\epsilon^2}{\sigma^2}\right)\\
		& = 1-\frac{\gamma(\frac{n-1}{2},\frac{n\epsilon^2}{2\sigma^2})}{\Gamma(\frac{n-1}{2})}
		\eea
		Here, $\Gamma(s)=\int_{0}^{\infty}t^{s-1}e^{-t}dt$, is the Gamma function and $\gamma(s,x)=\int_{0}^{x}t^{s-1}e^{-t}dt$ is the incomplete Gamma function. For the `bad bubbles' considered in the paper, we have
		\bea
		p(\alpha \log(L);\sigma,\epsilon) &= 1-\frac{\gamma\left(\frac{\alpha \log(L)-1}{2},\frac{\alpha \log(L)\epsilon^2}{2\sigma^2}\right)}{\Gamma\left(\frac{\alpha \log(L)-1}{2}\right)}\\
		&\approx 1-\frac{\gamma\left(\frac{\alpha \log(L)}{2},\frac{\alpha \log(L)\epsilon^2}{2\sigma^2}\right)}{\Gamma\left(\frac{\alpha \log(L)}{2}\right)}
		\eea
		Using the Chernoff bound we can bound this probability as
		\begin{equation}
		p(\alpha \log(L);\sigma,\epsilon) \leq \beta^{\alpha \log(L)},
		\end{equation}
		where $\beta=\left(\frac{\epsilon^2}{\sigma^2}e^{1-\frac{\epsilon^2}{\sigma^2}}\right)^{1/2}$.
		Hence, the probability $q(\alpha;\sigma,\epsilon)$ that there is no `bad bubble' in a length $L$ chain satisfies,
		\bea
		q(\alpha;\sigma,\epsilon) &\geq \lim\limits_{L\to \infty}\left(1-\beta^{\alpha \log(L)}\right)^{\frac{L}{\alpha \log(L)}}\\
		\log q(\alpha;\sigma,\epsilon) &\geq \lim\limits_{L\to \infty} \frac{L}{\alpha \log(L)} \log\left(1-\beta^{\alpha \log(L)}\right)
		\eea
		This proves Eq.7 in the paper (where we were concerned with the specific choice $\alpha=1/\zeta$). Since $q$ is a probability, $\log(q)\in(-\infty,0]$. The above bound is thus tight when the right hand side is $0$. The prefactor $L/\alpha\log(L)\to \infty$ in the limit, so the right hand side can be zero only when $ (L/\alpha\log(L))\log\left(1-\beta^{\alpha \log(L)}\right) \to 0 $. Expanding the logarithm (which is justified as $\beta<1$ for $\epsilon>\sigma$), we obtain
		\begin{equation}
		1+\alpha\log(\beta) <0
		\end{equation} i.e., $\beta < e^{-\frac{1}{\alpha}}$.
		
		We now explain the meaning of $\zeta$ in our discussion of the rare region model. In the discussion so far, $q(\alpha;\sigma,\epsilon)$ is the probability of having no $\alpha \log(L)$ sized rare regions in our spin chain. From the definition of rare region (any region whose local disorder exceeds $\epsilon$), it is clear that $q$ is a cumulative probability, $q=\int_{\epsilon}^{\infty}d\epsilon^{\prime}f(\alpha;\sigma,\epsilon^{\prime})$, where $f(\alpha;\sigma,\epsilon^{\prime})d\epsilon^{\prime}$ is the probability that there exist no $\alpha \log(L)$ sized regions whose local disorder is exactly $\epsilon^{\prime}$. Corresponding to $\epsilon^{\prime}$, there will be a logarithmic light cone, with an inverse length scale $\zeta^{\prime}(\epsilon^{\prime})$. The averaged time for information propagation across a chain with one rare region then, is given by,
		\bea
		t &\sim (L-\alpha\log(L))/v_B+\int_{\epsilon}^{\infty}d\epsilon^{\prime}f(\alpha;\sigma,\epsilon^{\prime})e^{\zeta^{\prime}\alpha\log(L)}\\&\sim(L-\alpha\log(L))/v_B+e^{\zeta\alpha\log(L)}
		\eea 
		where we have defined $\zeta$ as an averaged length scale associated with a rare region. In the paper we have argued that  $\zeta\alpha<1$ corresponds to a ballistic phase, which, along with the earlier condition with respect to $\alpha$, gives, 
		
		\begin{equation}
		\frac{\epsilon^2}{\sigma^2}e^{1-\frac{\epsilon^2}{\sigma^2}}<e^{-2\zeta}
		\end{equation}
		This proves Eq.8 of the paper.
	\end{appendices}

	\bibliography{MBL_OTOC}

\end{document}